\documentclass[a4paper]{jpconf}
\usepackage{graphicx}
\usepackage{amssymb}
\usepackage{amsmath}
\usepackage{ dsfont}
\newcommand{\ee}{\end{equation}}
\newcommand{\bb}{\begin{equation}}
\newcommand{\eqb}{\begin{eqnarray}}
\newcommand{\eqf}{\end{eqnarray}}

\begin{document}
\title{\vspace{-0.1cm}{\small \hfill{{DESY 11-170 }}}\\[1.8cm]Illuminating WISPs with photons}
\author{Paola Arias$^{a,b}$  and Andreas Ringwald$^{a}$}

\address{\em $^a$Deutsches Elektronen-Synchrotron, Notkestra\ss e 85, D-22607 Hamburg, Germany \\
 $^b$ Facultad de F\'isica, Pontificia Universidad Cat\'olica de Chile, Casilla 306, Santiago 22, Chile}
\ead{paola.arias@desy.de, andreas.ringwald@desy.de}

\begin{abstract}
Physics beyond the Standard Model naturally gives rise to very light and weakly interacting particles, dubbed WISPs (Weakly Interacting Slim Particles). A prime example is the axion, that has eluded experimental detection for more than thirty years. In this talk we will review some of the strongly motivated candidates for such particles, the observational hints for them and the present status of searches with photon regeneration experiments, as well as possible future improvements.
\end{abstract}
\section{Introduction}
Despite the success of the Standard Model of elementary particle physics (SM) there are still several missing ingredients for a successful description of the universe, the most prominent a dark matter candidate. Several proposals have been made to embed the SM in a more general and consistent unifying theory, evoking new physics. We have learned that a whole new group of very weakly and slim particles (WISPs) may emerge as a consequence of physics beyond the Standard Model.  An example among these particles is the axion, proposed to solve the strong CP problem \cite{Peccei:1977hh} of Quantum Chromodynamics (QCD), which has eluded detection till these days. On the other hand, unifying frameworks such as string theory predict the existence of axion-like particles (ALPs) \cite{Witten:1984dg}, i.e. particles that also emerge from the rupture of global symmetries, with the same interactions as the QCD axion but with different mass and decay constant. Another important candidate for a WISP is the so called hidden photon, a light extra $U(1)$ gauge boson \cite{Okun:1982xi} emerging from hidden sectors, commonly needed to break supersymmetry.
Generally, as a common feature, the weakness of the WISPs couplings to SM particles and the smallness of the mass are inherently related to a high energy scale at which the breaking of an underlying symmetry occurs. So, not only it is very plausible to have light, very weakly coupled particles, but indeed if we find them we may obtain information on the physics beyond the SM at very high energy scales.
The possible masses and couplings for WISPs are spanned over a wide range in parameter space, whose different regions are probed with astrophysical, cosmological and terrestrial searches in a complementary manner.  
Therefore astrophysical, and terrestrial searches are fundamental to constrain their existence. A notable difference to the case of weakly interacting massive particles (WIMPs)  is, however, that powerful 
accelerators  are not useful to detect WISPs, but 
instead powerful lasers and electromagnetic fields which allow for low energy experiments of high precision  \cite{Jaeckel:2010ni}.

The present talk is organized as follows: in section~\ref{sec1} we review the physics case for WISPs, only focusing on axions, ALPs and hidden photons. Several other candidates have been proposed \cite{chameleon}, but due to the high constraints on their existence we will not discuss them here, even though most of our analysis applies as well. In section~\ref{sec2} we briefly recall the present limits on WISPs set by cosmology and astrophysics. In section~\ref{sec3} we highlight puzzling observations, that could be explained by the existence of WISPs. In section~\ref{sec4} we introduce one of the most popular searches for WISPs that exploit the 
coupling to photons: laser regeneration experiments, also known as light shining through walls experiments (LSW) and we compute the conversion probability. In section~\ref{sec5} we recall some important improvements that have been proposed for LSW experiments and we also comment on their limitations. Finally, in section~\ref{sec6} we conclude showing the expected sensitivity of these experiments on axions, ALPs and hidden photons. 
\section{Physics case for WISPs}{\label{sec1}}
\subsection{Axion and axion-like particles}
As already noted, the axion was proposed as a way to solve the strong CP problem. Due to the Adler-Bell-Jackiw (ABJ) anomaly \cite{jackiw}, a CP-violating term should appear in the Lagrangian of strong interactions
\bb \mathcal L_{\rm QCD}\supset \frac{\alpha_s}{4\pi} \bar \theta G_{\mu\nu} \tilde G^{\mu \nu}, \, \, \, \, \,  \tilde G_{\mu\nu}=\epsilon_{\mu\nu\rho\lambda} G^{\rho\lambda}. \label{strong}\ee Where $G$ is the gluonic field strength, and $\alpha_s$ is the strong coupling constant. The parameter $\bar \theta=\theta +\mbox{arg det} M$ is the sum of the CP violating term arising from the ABJ anomaly and the argument of the determinant of the complex quark matrix. This parameter is not constrained by the theory and must be determined experimentally. A sensitive probe of $\theta$ is provided by the measurement of the electric dipole moment of the neutron, which would emerge from a term such as equation~(\ref{strong}). The theoretical computation predicts $|d_n|\sim 10^{-16}\bar \theta $ e cm, and from the recent experimental bound \cite{Baker:2006ts} on $|d_n|<2.9\times 10^{-26}$ cm e, it is possible to set the limit of $\bar \theta  < 10^{-10}$, a really small number. The strong CP problem is the puzzle why the sum of two unrelated quantities is so unnaturally small.

To solve this problem, a new chiral $U(1)$ symmetry is introduced in the SM- the so called Peccei-Quinn (PQ) symmetry \cite{Peccei:1977hh}, which is spontaneously broken. The pseudo-Goldston boson associated is the axion $a$, which has a shift symmetry only broken by the CP-violating term 
\bb \mathcal L_a= -\frac{1}2 \left(\partial_{\mu}a\right)^2+\frac{a}{f_a}\frac{\alpha_s}{4\pi}G_{\mu\nu} \tilde G^{\mu \nu} +\mathcal L\left(\partial a;\psi\right).\ee  This means that the axion field has a non-zero potential, and therefore has a vacuum expectation value different from zero, given by $\langle a\rangle=-\bar \theta f_a$. Thus, the $\bar \theta$ CP violating term can be absorbed into the axion field, defining the physical axion field as $a_{\rm phys}=a-\bar \theta f_a$.

Therefore, the introduction of the spontaneously broken $U(1)_{PQ}$ solves the strong CP problem, with the price of a new scalar particle, so far undetected. The axion is nominally massless as a Goldston boson, however acquires a small mass as a result of the chiral anomaly, namely
\bb m_a^2= \langle \frac{\partial^2 V_{a}}{\partial a^2}\rangle=-\frac{\alpha_s}{4\pi f_a} \frac{\partial}{\partial a} \langle G_{\mu\nu}\tilde G^{\mu \nu} \rangle|_{\langle a\rangle}.\ee
The axion mass can be expressed in terms of the light ($u,d$) quark masses, the pion mass $m_{\pi}$ and the pion decay constant $f_{\pi}$ as \cite{Nakamura:2010zzi}:
\bb m_a=\frac{m_{\pi}f_{\pi}}{f_a}\frac{\sqrt{m_u m_d}}{m_u+m_d}=\frac{0.60 \, \mbox{meV}}{f_a/10^{10}\,  \mbox{GeV}}. \ee
There are two benchmark invisible ($f_a \gg v_{\rm weak}$) axion models. The model known as KSVZ \cite{ksvz} considers new heavy quarks carrying $U(1)_{\rm PQ}$ charges, leaving normal quarks and leptons without tree-level couplings. In models known as DFSZ \cite{dfsz} at least two Higgs doublets are needed and ordinary quarks and leptons carry PQ charges. The coupling of axions to two photons 
\bb \mathcal L_{a\gamma\gamma}=-\frac{1}4 g_{a\gamma\gamma}a F_{\mu\nu} \tilde F^{\mu\nu}=g_{a\gamma\gamma}\, a\, \vec{E}\cdot \vec{B},\ee
is very important for many experimental searches. The coupling constant $g_{a\gamma\gamma}$ is model dependent and is given by
\bb g_{a\gamma\gamma}=\frac{\alpha}{2\pi f_a}\left(\frac{E}{N}-\frac{2}3\frac{4+z}{1+z} \right)\sim 10^{-13}\, \mbox{GeV}^{-1} \left(\frac{10^{10}\mbox{GeV}}{f_a}\right), \ee where $z=m_u/m_d$ and $E$ and $N$ are the electromagnetic and color anomalies associated with the axion anomaly. For KSVZ models, $E/N=0$, and for DFSZ models, $E/N=8/3$.

The concept of an extra $U(1)$ symmetry that is spontaneously broken has been generalized to give rise to particles that may share the same coupling as axions (the most relevant being the coupling to photons), but with a totally different origin. These particles have therefore been dubbed axion-like particles (ALPs) and they can be found in a much more wider region in parameter space, since there is no a priori relation between the mass and the coupling constant, such as for the axion. An example of ALP would be a generic pseudo-Nambu-Goldstone-Boson emerging as particle excitations of fields that acquire some vacuum expectation value, due to a spontaneous symmetry breaking. In this case the smallness of the ALPs masses is inversely related to the very high energy scale of new physics. For instance if the symmetry is a $U(1)$ chiral symmetry, it would be very likely to find a coupling to two photons 
\bb \mathcal L \supset g\phi F_{\mu\nu} \tilde F^{\mu\nu}. \ee
Besides the pseudo-Nambu-Goldsone fields it is possible to find couplings of  new scalar bosons  to photons such as the equation above from string compactifications where the generically present zero modes of antisymmetric tensor fields coupled to gauge fields via Chern-Simons terms lead to CP violating couplings in the low energy effective theory \cite{Witten:1984dg}.
\subsection{Ultralight Hidden-Sector Particles}
Another well motivated WISP are  the hidden sector $U(1)$ gauge bosons, or hidden photons. They are a generic feature arising from string compactifications. Usually hidden sectors are only weakly coupled to the visible sector via gravitational interactions and after compactifications, their gauge groups may have broken into products of non-Abelian groups and $U(1)$ gauge groups. Observable effects are strongly suppressed because interactions occur through operators of mass dimension greater than $n=$4; at low energies they go as $\left( E/M_s\right)^{n-4}$, where $E$ is the effective low energy scale and $M_s$ is the string scale.
However, remarkable exceptions are hidden sector Abelian gauge bosons - messengers between the hidden and visible sectors - whose $U(1)$ may remain unbroken down to very small energy scales. The dominant interaction of the hidden photon with the SM particles is with photons, through a kinetic mixing term
\bb \mathcal L \supset -\frac{1}{4}F_{\mu\nu} F^{\mu\nu} -\frac{1}{4}X_{\mu\nu} X^{\mu\nu} +\frac{\chi}{2} F_{\mu \nu} X^{\mu\nu} +\frac{m_{\gamma'}^2}{2 } X_\mu X^\mu, \label{hiddphoton} \ee where $X_\mu$ denotes the hidden photon field, with field strength $X_{\mu\nu}$. The strength of the mixing with photons is encoded in the dimensionless parameter $\chi$ generated at loop level via heavy messenger exchange, predicted to be very small. Its value usually\footnote{In the context of compactifications of the heterotic string, the size of the mixing can be in the range of $10^{-5}\div10^{-17}$ \cite{Dienes:1996zr, Goodsell:2010ie}} fluctuates between \cite{Abel:2008ai, Goodsell:2009xc, Cicoli:2011yh} \bb 10^{-12} \lesssim \chi \lesssim 10^{-3}.\ee
We have also included a mass term for the hidden photon in the effective Lagrangian equation~(\ref{hiddphoton}) arising from a standard Higgs mechanism or a Stueckelberg mechanism. In the latter case, the mass and the size of the kinetic mixing are typically linked through the string scale as
\bb m_{\gamma'}^2|_{\rm stuck} \eqsim \frac{g_s}2 \left(\frac{4\pi}{g_s^2}\frac{M_s^2}{M_P^2}\right)^{z}, \, \, \, \, \,  z=\frac{1}3, 1.\ee Where $g_s$ is and $M_P$ is the Planck mass. Therefore in the Stueckelberg mechanism case, the discovery of a hidden photon translates into a prediction of the string scale.

\section{Constraints on WISPs from Astrophysics and Cosmology}{\label{sec2}}
\begin{figure}[t]
\begin{minipage}{16pc}
\includegraphics[width=16pc]{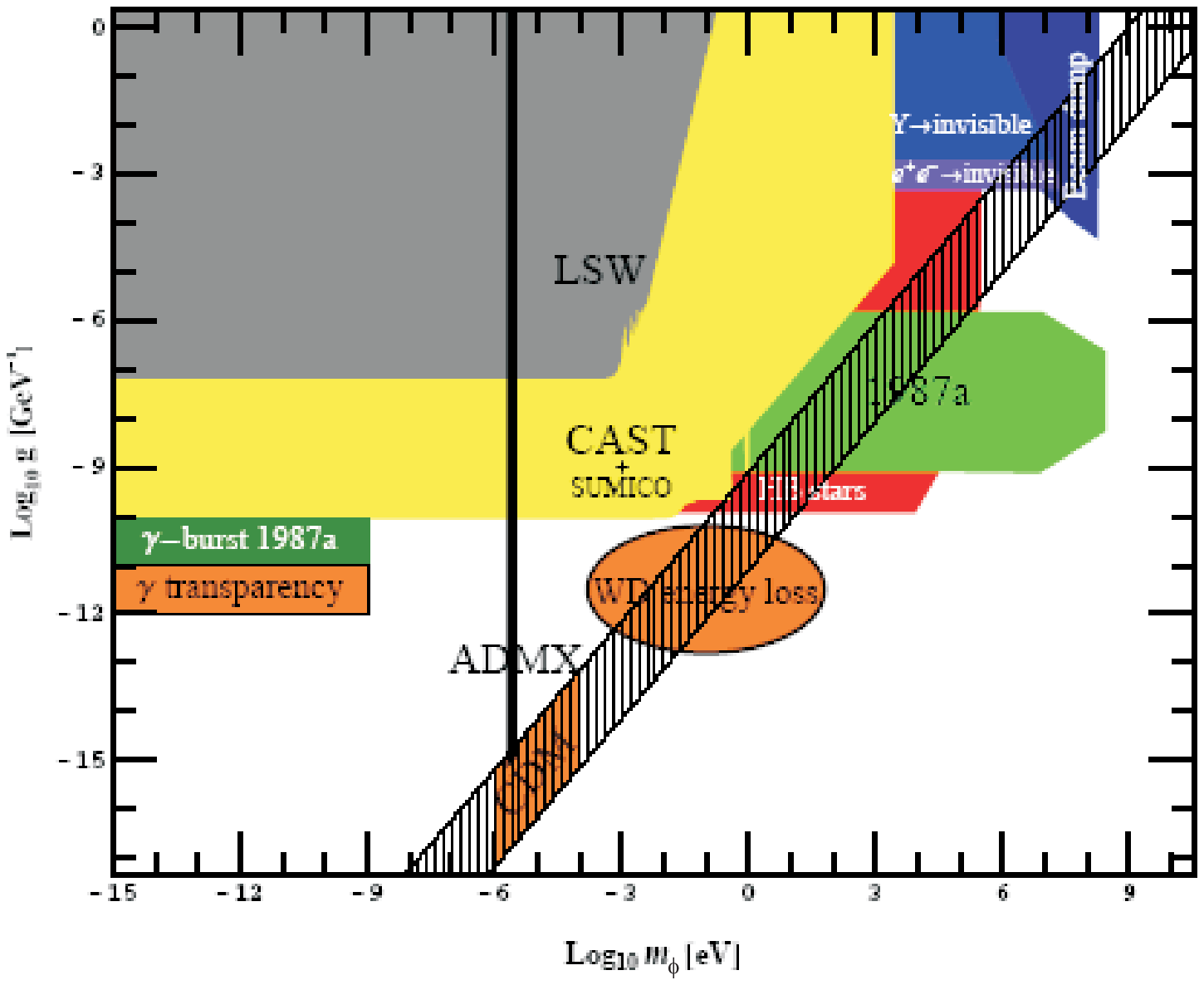}
\caption{\label{fig:constraints1} \scriptsize{Summary of cosmological and astrophysical constraints for axions and axion-like-particles (left). See the text for details. Orange circled regions correspond to hint on ALPs. Compilation from reference~\cite{Jaeckel:2010ni} where also more details can be found.}}
\end{minipage}\hspace{2pc}%
\begin{minipage}{20pc}
\includegraphics[width=20pc]{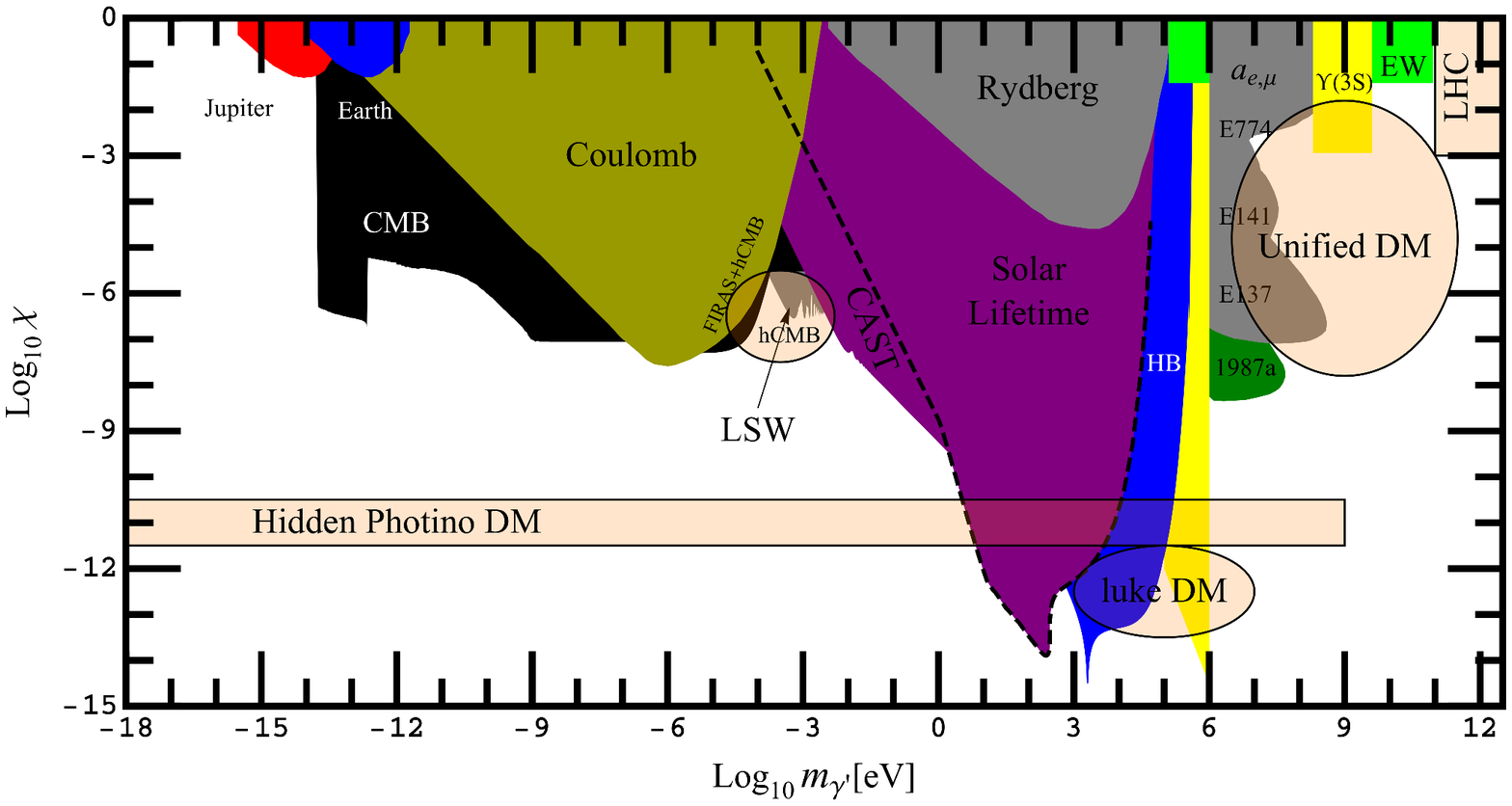}
\caption{\label{fig:constraints2} \scriptsize{ Summary of cosmological and astrophysical constraints for hidden photons. See the text for details.  Areas that are especially interesting are marked in light orange. Compilation from reference~\cite{Jaeckel:2010ni} where also more details can be found.}}
\end{minipage} 
\end{figure}

Several observational and experimental techniques are used to search for WISPs and they are summarized in figure~\ref{fig:constraints1} for axions and ALPs and in figure~\ref{fig:constraints2} for hidden photons. In principle it is not an easy task, since these particles can span a wide range in parameter space and their weak coupling to SM particles makes them really hard to detect. The most prominent and stringent  constraints come from astrophysics and cosmology, and we discuss them in this section
\subsection{Bounds from stellar evolution}
The emission of weakly interacting particles from stars usually leads to a modification of its evolutionary time scale \cite{Raffelt:1996wa}. For instance, the energy loss due to the new particle makes the burning of a star faster, diminishing the time of shining. However, this is not the case in low-mass red giants, where the emission of new particles would lead to a delay  of helium ignition, extending the red-giant phase. By identifying these evolutionary phases- sensitive to new energy losses channels - has been possible to set bounds on WISPs in different stellar environments \cite{Raffelt:1996wa, Raffelt:2006cw}. From figure~\ref{fig:constraints1} one can see that the most stringent bounds on axions are coming from Supernova 1987a and observation of horizontal branch stars (HB) stars. This is because in these objects the temperature is the adequate so neutrino and axion emission is important compared to photon emission, and neutrinos are just starting to have an impact on stellar observables, therefore axion emission is not eclipsed. On the other hand, the sun lifetime does not set a bound on axions nor ALPs but on hidden photons \cite{Redondo:2008aa} and is complemented  by the lifetime of HB stars \cite{Redondo:2008ec}.
\subsection{Bounds from big bang nucleosynthesis and the cosmic microwave background}
After the realization that the universe is expanding, several measurements of the expansion rate have enlightened the physics in the early universe.  Primordial abundances, cosmic microwave background (CMB) anisotropies and large scale structure allow us to infer the  particle content in the past. A sensitive measurement of the expansion rate of the universe during big bang nucleosynthesis (BBN) is the $^4H_e$ mass fraction, $Y_p$, which according to WMAP is $Y_p=0.2486 \pm 0.0002$ ($68 \% $ C.L.) \cite{Cyburt:2008kw} in the frame of SM cosmology. However, several estimates of the helium abundance seem to indicate a (yet not conclusive) excess 
of $Y_p$ which can be attributed to an extra degree of freedom, enclosed in a higher number of effective neutrinos \cite{Nakayama:2010vs}. Recent determinations of this number seem to favor an excess at the BBN epoch of  $\Delta N_\nu^{\rm eff} \sim 1$ and therefore can be used to motivate or constrain WISPs that may have added new relativistic degrees of freedom. As noted in \cite{Jaeckel:2010ni, Nakayama:2010vs} an extra neutral spin-0 particle thermalized during BBN is allowed, but it is not the case for a hidden photon.

Another sensible measurement on new light particles is the CMB anisotropy. Reactions involving $\gamma \rightarrow \rm{WISP}$ conversion would have depleted photons in a frequency-dependent way, that can be tested with the spectrum measurements by FIRAS \cite{Fixsen:1996nj}, and have been used to set constraints on light ALPs \cite{Melchiorri:2007sq} and hidden photons \cite{Jaeckel:2008fi}. Besides, it has been noted in \cite{Mirizzi:2009iz} that resonant production of hidden photons  would lead to a distortion in the CMB spectrum, providing a strong constraint on these particles.

\section{Hints on WISPs}{\label{sec3}}
In this section we will review some possible hints on the existence of WISPs, they have been marked in orange in the case of ALPs in figure~\ref{fig:constraints1} and circled in light yellow in figure~\ref{fig:constraints2}  for the case of hidden photons.
\subsection{Hints on ALPs}
One of the most important hints on ALPs is the possibility that they may comprise all or some part of the cold dark matter in the universe. This can be achieved by the so-called misalignment mechanism: the ALP should never reach thermal equilibrium, however the misalignment mechanism will excite coherent oscillations of the field. If it acquires a tiny mass the field will begin to move and eventually when the mass exceeds the expansion rate of the universe, will start to oscillate, populating the universe. At first this idea was brought up in the context of the axion \cite{Sikivie:2006ni} with masses in the range of the $m_a \sim \mu $eV, but it has been noted that the mechanism can also be applied very well to ALPs \cite{nos}, where a much wider parameter space region is allowed.
A second extra hint comes from the observation of a non standard energy loss in white dwarfs \cite{Isern:2008nt}, compatible with an ALP of $g_\phi\sim 10^{-12}$ GeV$^{-1}$ and a mass $m_{\phi}\lesssim$ meV.
Finally, the fact that distant astrophysical $\gamma$ sources have been observed in the range of the TeV by H.E.S.S and  MAGIC -- a fact which is puzzling 
since it is believed that $\gamma $ absorption from extragalactic background light is too strong to allow their observation \cite{Aharonian:2005gh} --- may be attributed to $\gamma \to$ ALP oscillations. Indeed, such conversion could take place in the magnetic fields around the $\gamma$ sources, allowing the ALP to travel undamped till they reach our galaxy, where the back conversion may take place in the intergalactic magnetic field \cite{DeAngelis:2007dy, Hooper:2007bq}, although this is highly dependent on the strength and location of the fields. The apparent transparency of the universe to gamma rays favors an ALP of mass $m_\phi \lesssim 10^{-9}$~eV and a coupling constant of $g_\phi \sim 10^{-12}$~GeV$^{-1}$, excluding an axion interpretation, but instead an axion-like particle.

\subsection{Hints on hidden photons}
Resonant oscillations between photons and hidden photons after BBN and before decoupling may comprise a hidden CMB \cite{Jaeckel:2008fi}. The hidden photons in the range of $m_{{\gamma'}}\sim$~meV and $\chi \sim 10^{{-6}}$  produced by such oscillations  would constitute a contribution to the dark radiation at the CMB epoch, leading to an
apparent increase of the effective number of neutrinos. Interestingly, according to several recent observations, a number higher than three is currently favored \cite{Komatsu:2010fb}.  This observation will soon be tested by the PLANCK satelite. At the same time, a hidden photon in the parameter range of 
interest can also be searched for by the  LSW experiment ALPS (see below).

Another important hint for the existence of hidden photons is their relation with dark matter. As mentioned before, the misalignment mechanism will lead to cold dark matter production 
for any light conditions. The latter may 
be fulfilled by hidden photons \cite{Nelson:2011sf}. Resonant Compton evaporation of hidden photons is the main process that could thermalize it, and thus  constrains the region where could be found as dark matter \cite{nos}.  The favored region is quite wide, $(\chi, m_{\gamma'})\lesssim(10^{-9}, \mbox{eV})$, and very encouraging for laboratory experiments that could test it in the near future. Other possibilities are  that the hidden photon may be a lukewarm dark matter 
candidate in the range of $(\chi, m_{{\gamma'}}) \sim (10^{{-12}},0.1\mbox{MeV})$ \cite{Redondo:2008ec}, and that a heavy hidden photon in the range of $(\chi, m_{{\gamma'}}) \sim (10^{-4}, \mbox{GeV})$ could play an important role in models where the dark matter resides in the hidden sector \cite{ArkaniHamed:2008qn, Pospelov:2008jd, Andreas:2011in}.

\section{Laboratory searches with photons}{\label{sec4}}

Several dedicated experiments are looking for WISPs worldwide, and they have contributed to constrain the parameter space on these particles. Laboratory searches are in many cases not competitive to 
astrophysical searches, e.g. with 
helioscopes  \cite{Andriamonje:2009dx}, but still they help to provide important confirmation on these searches, due to their clean and controlled environment. One of the most promising laboratory searches are the so-called light shining through a wall (LSW) experiments, which exploit one of the most attractive features of WISPs, their oscillation with photons. We will proceed to review these experiments briefly in the next subsection (for an exhaustive review, see reference \cite{Redondo:2010dp}).

\subsection{Photon $\rightarrow$ WISP oscillations} 
Oscillations between photons and WISPs (and viceversa) are possible due to an effective non diagonal "mass matrix" , $\mathcal M$, that arises due to the couplings between them. The generic equation of motion looks like \cite{Jaeckel:2010ni,Raffelt:1987im}
\bb \left[ \left(\omega^2 +\partial_z^2\right) \mathds{1} - \mathcal M \right] \mathcal V=0, \ee where $\mathcal V$ is a composite vector between photon components and the WISP field (generally the different photon polarizations have different equations of motion). The generic solution to these equations can be found by diagonalization of the mass matrix $\mathcal M$ and writing the interaction states $\mathcal V$ as a combination of the propagation eigenstates of the system $ \mathcal U$, which are given by

\bb\mathcal U= D^\dag \mathcal V, \ee where $D$ is the matrix that diagonalizes $\mathcal M$. The mass of the propagation states is given by the eigenvalues of the mass matrix, and therefore, the wave number for each propagation eigenstate is $k_{1,2}^2=\omega^2 -\lambda_{1,2}$, where $\lambda_{1,2}$ are the two eigenvalues of $\mathcal M$. The details of the mass matrix depend on the type of boson that we are considering. For axions and ALPs the interaction occurs only with the component of the photon parallel to the external magnetic field $B$, and the mixing matrix is given by
\bb \mathcal M_{\phi}= \begin{pmatrix} 0 & g\, B\, \omega \\
g\, B\, \omega &  m_{\phi}^2\end{pmatrix}.\ee Meanwhile, the mass matrix for hidden photons is given by

\bb \mathcal M_{\gamma'}= m_{\gamma'}^2\begin{pmatrix} \chi^2 & \chi \\
\chi &  1\end{pmatrix},\ee 
The probability that a photon oscillates into a WISP\footnote{Recall that in this review we are only focusing on axions, ALPs and hidden photons as WISPs.} after a traveling distance $L$ reads

\bb P_{\gamma\rightarrow \mbox{WISP}}= 4 \sin^22\theta \sin^2\left(\frac{|k_\gamma-k_\phi| L}{2}\right), \label{prob} \ee  where the momentum transfer between photon and WISP is given by $|k_\gamma-k_\phi|= |\omega-\sqrt{\omega^2-m_{\rm WISP}^2} |\approx m^2_{\rm WISP}/\left(2\omega\right)$, and the oscillation angle satisfies 
\bb \tan2\theta=\frac{\mathcal M_{12}}{\mathcal M_{11}-\mathcal M_{22}}.\ee
The corresponding mixing angles for ALPs and hidden photons are given by
\bb \sin^22\theta_{\phi} = \frac{g^2 B^2 \omega^2}{m_\phi^4},  \, \, \, \, \, \, \, \,  \sin^22\theta_{\gamma'}= \chi^2.\ee
\subsection{Light shining through a wall experiments}
\begin{figure}
\begin{center}
\includegraphics[scale=0.6]{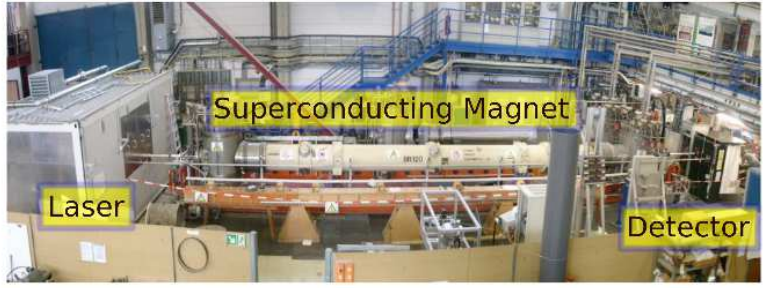}
\end{center}
\caption{\label{fig:LSW}\scriptsize{Schematic of a LSW setup (ALPS experiment at DESY). The laser is injected in the bore of the superconducting dipole magnet. An opaque wall is placed at the center of the magnet. WISPs produced in this part, called production side pass through the wall, to the regeneration side where can reconvert into light and continue to the detector in the end part of the experiment. Figure from reference~\cite{Redondo:2010dp}.}}
\end{figure}
In a  light shining through a wall experiment the light shines into an opaque wall (magnetic region is needed for the case of ALPs) and if the conversion $\gamma \rightarrow \mbox{WISP}$ took place, the latter will be able to pass through the wall. On the other side, an exact same region makes possible the reconversion into a photon that can reach the detector, making possible the phenomenon of light shining through a wall. The schematic of the experiment is depicted in figure~\ref{fig:LSW}. Since the photon must be regenerated on the other side of the wall, the total probability to detect it in a symmetric setup is
\bb P_{\rm LSW}= P_{\gamma\rightarrow \mbox{WISP}}^2.\ee
As can be inferred, the product $B\,L$ is the most straightforward way to increase the sensitivity of the experiment for an axion or an ALP. Another important feature of LSW experiments is the use of lasers in the optical regime, since they provide the highest photon fluxes in order to overcome the smallness of the probability of conversion.
In the case of ALP searches, LSW experiments are not yet competitive with other solar or astrophysical searches, (see figure~\ref{fig:constraints1}) but they are exploring new parameter space in the case of hidden photons (see figure~\ref{fig:constraints2}) and they have a huge potential to increase their sensitivity due to several optimization techniques that have been already proposed but are still not implemented. In the most recent generation of laser regeneration experiments a sensitivity of the order of $g \lesssim 10^{-8}$ GeV$^{-1}$ has been reached for ALPs and $\chi \lesssim 10^{-6}$ in the case of hidden photons.
A total of six LSW experiments have been performed all over the world, in alphabetical order they are: ALPS  at DESY \cite{Ehret:2010mh} with a magnetic strength of $B=5.5$~T, BFRT at Brookhaven \cite{Ruoso:1992nx} with $B=3.7$~T, BMV at LULI  \cite{Robilliard:2007bq} with $B=12.3$~T, GammeV at Fermilab \cite{Chou:2007zzc} with $B=5$~T, LIPSS at JLAB~\cite{Afanasev:2008fv} with $B=1.7$~T and finally OSQAR at CERN \cite{Pugnat:2007nu} with $B=9$~T. Many of them are already planning and implementing a next phase.
\section{Optimizing light shining through a wall experiments}{\label{sec5}}
For the next generation of LSW experiments several important proposals have already been made to increase the sensitivity. One of the most
promising techniques is to include high quality matched Fabry-Perot cavities in the production and regeneration sides of the
experiment~\cite{Hoogeveen:1990vq}. When both cavities are tuned to the same frequency, $\omega$, it is possible to gain an enhancement
in the sensitivity for the ALP-photon coupling or the kinetic mixing parameter by the fourth root of each of the
cavities' power buildups\footnote{We chose to refer to the power buildup of the cavity, $\beta $, instead of the commonly used finesse {since it is the former which plays the most direct role in the production and regeneration of WISPs. }}, i.e $(g, \chi)  \propto (\beta_g \beta_r)^{-1/4}$.
Considering that with the available technology cavities with $\beta \sim 10^4-10^5 $ seem realistic, an improvement of the order of $10^2$ in these couplings are feasible.
The expected number of photons after the regeneration cavity of such LSW experiment will be given by~{\cite{Hoogeveen:1990vq,Mueller:2009wt}}
\bb
N_s=\eta^2 \, \beta_g \beta_r \, \frac{\mathcal{P}_{\rm prim}}{\omega}\, 
P_{\gamma\rightarrow {\rm WISP}}^2 \tau,
\label{regphoton}
\ee
where $\mathcal{P}_{\rm prim}$ is the primary laser power,  $\beta_{g,r}$ are the power build-ups of the generation and regeneration cavities,
$\eta$ is the spatial overlap integral between the WISP mode and the electric field mode~\cite{Hoogeveen:1990vq} and $\tau$ is the measurement time.
Most likely a second improvement for the next generation of LSW experiments will be to enlarge the magnetic region, without losing magnetic strength. This can be achieved by arranging several dipole magnets in a row \cite{Ringwald:2003nsa}. Assuming a setup with $6+6$ HERA dipole magnets, the expected sensitivity will be of the order of
 \eqb
 \nonumber
 g_{\rm sens} &=&\frac{ {2.71} \times 10^{-11}}{\mbox{GeV}} 
 \left[\frac{290 \ \mbox{Tm}}{BL}\right]
 \left[\frac{0.95}\eta\right]^{1/2}							\\
&& \times 
 \left[\frac{10^{10}}{\beta_g\beta_r}\right]^{1/4}
\left[ \frac{3 \ {\rm W}} {{\mathcal P}_{\rm prim}}\right]^{1/4}	
\left[\frac{n_b}{10^{-4} {\rm Hz}}\right]^{1/8}
\left[\frac{100 {\rm h}}{\tau}\right]^{1/8},
 \label{gmin}
 \eqf
 \begin{table}[t]
\caption{\label{table:t1}{\scriptsize{Benchmark values for a next generation LSW experiment.}}}
\begin{center}
\begin{tabular}{llllllll} 
\br 
$6+6$ HERA  & B & ${\mathcal P}_{\rm prim} $ & $\beta$ & $\omega$ & $\eta$ & $\tau$&$n_b$\\
\mr
$L= 52.8$~m & $ 5.5$~T 
& 3~W  & $\beta_g=\beta_r=10^{5}$ & $1.17$~eV  & $ 0.95$ & $100$~h & $ 10^{-4}$~Hz \\
\br
\end{tabular}
\end{center}
\end{table}%
where we have used the benchmark values for the most important parameters, 
as summarized in table~\ref{table:t1}, and we have been quite conservative, assuming that no single photon detection could be achieved, therefore we have included the dark count rate, $n_b$. However, care must be taken with  respect to two important points: first, the enlargement of the resonant cavity by adding $N$ magnets is strongly dependent on the diameter of the laser beam and therefore, the aperture of the magnet, and secondly when arranging several magnets in a row, we are also including a natural and probably unavoidable ÒgapÓ, with no magnetic field in between each magnet. The impact of this gap has been proven to be non-negligible \cite{Arias:2010bh} and in order to better optimize the experiment should be taken into account. Of course the second point only applies to ALP searches with LSW experiments, since for hidden photons the magnetic region makes no difference.
Let us start by addressing the gap, $\Delta$, in between the magnets. The probability to detect a photon after the wall gets modified whether the arrangement of magnets is the normal one (all the same polarization) or a wiggler (alternate polarization \cite{VanBibber:1987rq}), for both cases they are given by
\eqb
P_{\gamma\rightarrow\phi}=\frac{1}4\frac{\omega}{k_\phi}\left(gBL\right)^2
\left|F(qL)\right|^{2},
\label{prob}
\eqf with $q=|k_\gamma-k_\phi| $ and $F(qL)$ a function known as the form factor. In the case of just one magnet, takes the form $F(L)=\frac{2q}L \sin\left(\frac{qL}2\right)$. For an arrangement of $N$ magnets of length $\ell$ each, and the same polarization the form factor is given by \cite{Arias:2010bh}
\eqb 
\left|F_{N,\Delta}(qL)\right|
= \left|\ \frac{2}{qL}\sin\left(\frac{qL}{2N}\right)\frac{\sin\left(\frac{qN}{2}\left(\frac{L}N+\Delta\right)\right)}
{\sin\left(\frac{q}{2}\left(\frac{L}N+\Delta\right)\right)}\right|,
\label{fgaps}
\eqf%
and in the case of a wiggler configuration of $n$ alternating regions, it takes the form
\bb \left|F_n (qL)\right| = \left\{
\begin{array}{l l}
\left|\frac{2}{qL}\sin\left(\frac{qL}{2}\right)\tan\left(\frac{qL}{2n}\right)\right|{,} & \quad \text{$n$ even}{,}\\
\left|\frac{2}{qL}\cos\left(\frac{qL}{2}\right)\tan\left(\frac{qL}{2n}\right)\right|{,} & \quad \text{$n$ odd}{.}\\
\end{array} \right. \label{wiggler}\ee%
Where in both cases, the total magnetic length is $L=N\ell$. We can now maximize equations~(\ref{fgaps}) or~(\ref{wiggler}) varying the gap length. Thus, we are able Ð in principle Ð to optimize the sensitivity for given values of $m_{\phi}$ changing the size of the gap, scanning optimally the ALP parameter space. For instance, maximization of equation~(\ref{fgaps}) gives $(q\ell/2) (1 + \Delta_{\rm opt} /\ell) = k\phi$, with $ k \in \mathds{Z}$. Unfortunately, using this equation, a full scan of the parameter region it is not possible experimentally, because Æ is limited by the length of the setup, and in particular also by the maximal length of the cavity (see below).

As we mentioned before, attached to the enlargement of the magnetic region, comes the issue of the optimal length of the Fabry-Perot cavity. On the one hand, the cavity should be as large as possible, since in the case of ALPs, the conversion probability is directly proportional to the square of the total length, and in the case of hidden photons, large cavities allow to have good sensitivity at lower masses. But on the other hand, large cavities may introduce high diffraction losses. This leads to minimum requirements on the diameter of the laser beam and therefore on the aperture of the cavity, which of course, is set by the aperture of the magnet.
In order to achieve a high power build-up of the cavity (as we have assumed in our estimations before) the losses should be kept to a minimum. In particular, for an impedance matched cavity the most important losses come from the transmissivity of the cavity mirrors and the round trip losses, specially from clipping the cavity mode. Following \cite{Arias:2010bh} the sensitive coupling constant of photons to ALPs (for hidden photons it is exactly the same) depends mainly on the combination
\bb
\label{opti}
 g_{\rm sens}\propto \frac{1}{L{\beta^{1/2}}}= \frac{1}{L}\left(e^{-2 a^2/w^2(Z)}+{\delta_0^*}+\delta_2\right)^{1/2}, 
\ee where the exponential factor accounts for the clipping losses, and $a$ accounts for the aperture of the cavity (set by the aperture of the magnet), $w(Z)$ is the spot of the laser beam at some distance $Z$ from the source,  $\delta_2$ for the losses on the transmissivity of the mirrors and $\delta_0^*$ accounts for other contributions to internal losses of the cavity.  Therefore, the optimization of the cavity length comes to keep the coupling constant ($g, \chi$) as small as possible, including all these loss factors. Performing the minimization of equation~(\ref{opti}) it is possible to find that the optimal length of the cavity, $Z_{\rm opt}$, that ensures the best relation length-power build-up is given by
\bb
\label{miniformula}
Z_{\rm opt}=0.0755 \frac{\pi a^2}{\lambda}=89.2\ {\rm m} \left(\frac{a}{20\ {\rm mm}}\right)^2 \frac{1064\ {\rm nm}}{\lambda},  
\ee 
where $\lambda$ is the wavelength of the laser and it can be checked that this result {corresponds to ${\delta_0^{\rm clip}/(\delta_0+\delta_2)}=0.177$.
In previous literature~\cite{Mueller:2009wt} a value ${\delta_0^{\rm clip}/(\delta_0+\delta_2)}\sim 1$ was used, leading to a slightly less optimal setup. The optimal relation between the coupling constant and the number of magnets for the three currently available dipole magnets HERA, LHC and Tevatron can be 
inferred from the minima in figure~\ref{fig:opt1}. In figure~\ref{fig:opt2} the influence of the gaps between the magnets in the sensitivity of the experiment is exposed, assuming different gaps of $\Delta=1,2,3,4,5$ m. 
\begin{figure}[t]
\centering
\begin{minipage}{16pc}
\includegraphics[width=15pc]{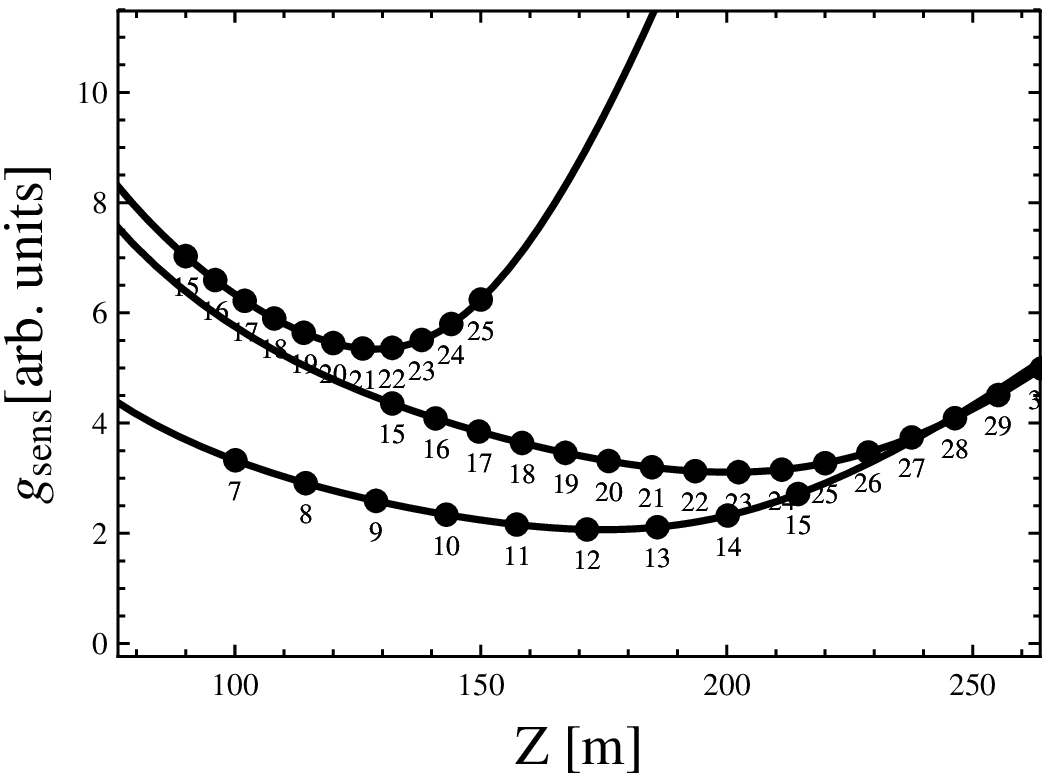}
\caption{\label{fig:opt1} \scriptsize{Comparison of $g_{\rm sens}$ vs. cavity length $Z$, between three possible LSW setups with HERA, LHC, and Tevatron magnets, in the ideal scenario of no gap in-between the magnets. The radius of the bore aperture of the different magnets is approximately $30$~mm for HERA dipoles, 28~mm for LHC magnets and 24~mm for Tevatron dipole magnets. Figure from reference~\cite{Arias:2010bh} .}}
\end{minipage}\hspace{2pc}%
\begin{minipage}{16pc}
\includegraphics[width=15pc]{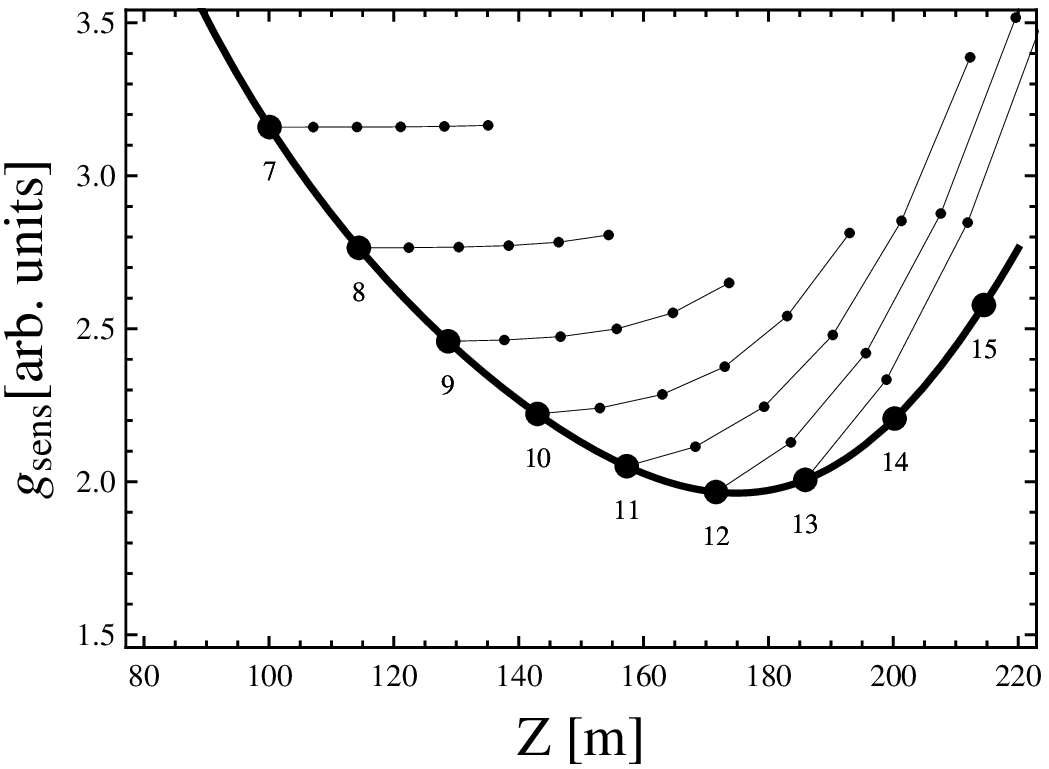}
\caption{\label{fig:opt2} \scriptsize{Quantifying the effect of the gap in the setup. The thin lines emerging from each gapless configuration to the right and up represent configurations with gap size $\Delta=1,2,3,4,5$ m. Figure from reference~\cite{Arias:2010bh} where also more details can be found. }}
\end{minipage} 
\end{figure}
\section{Sensitivity of the next generation of LSW experiments and outlook}{\label{sec6}}
\begin{figure}[t]
\centering
\begin{minipage}{16pc}
\includegraphics[width=15pc]{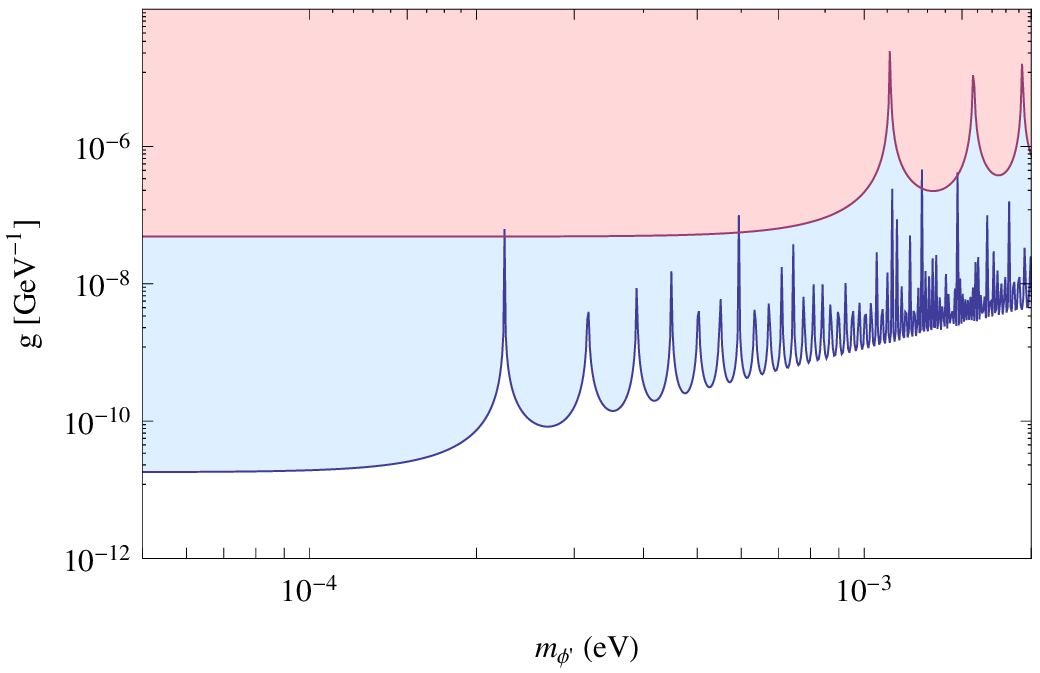}
\caption{\label{fig:bound1} \scriptsize{ALPs searches with light shining through a wall experiments. The red region corresponds to the best sensitivity of the previous phase of photon regeneration experiments (ALPS results). In light blue it is shown the discovery potential of the next generation. We have used the benchmark values of table~\ref{table:t1}. }}
\end{minipage}\hspace{2pc}%
\begin{minipage}{16pc}
\includegraphics[width=15pc]{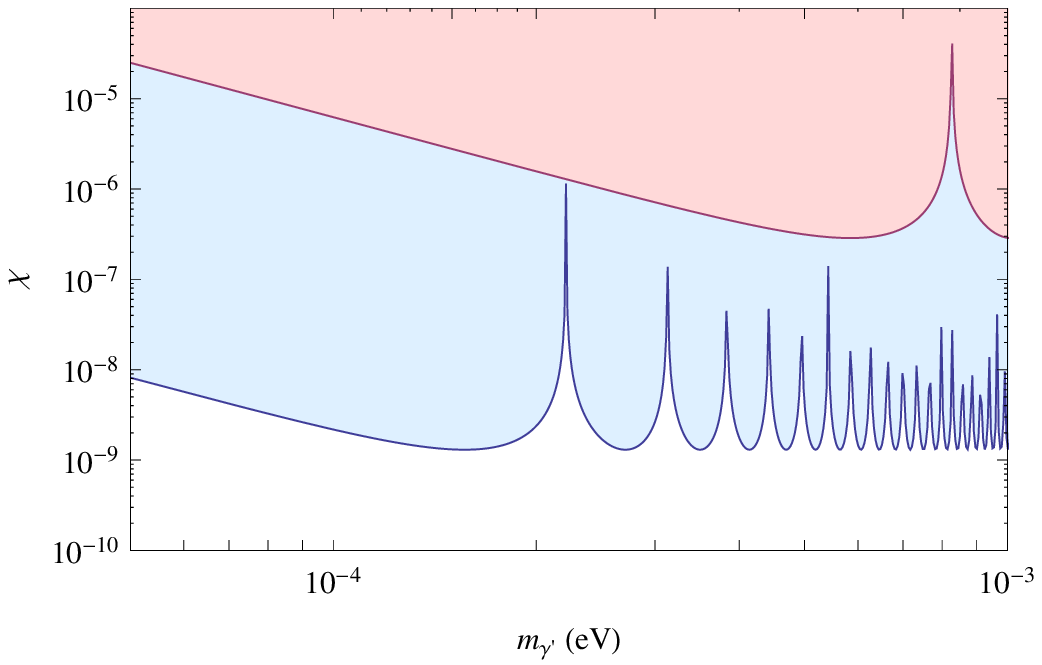}
\caption{\label{fig:bound2} \scriptsize{Hidden photon searches with light shining through a wall experiments. The red region corresponds to the best sensitivity of the previous phase of photon regeneration experiments (ALPS results). In light blue it is shown the discovery potential of the next generation. We have used the benchmark values of table~\ref{table:t1}. }}
\end{minipage} 
\end{figure}
The previous generation of LSW experiments have exploited just one or two superconducting dipole magnets. Nonetheless, the most sensitive experiment so far is the Any Light Particle Seach (ALPS) experiment \cite{Ehret:2010mh} that used only one HERA dipole. They were able to incorporate mirrors to enhance the conversion probability in the generation side, reaching a power build-up of the cavity of around $\beta_p\sim 300$. Another important achievement of the collaboration was to successfully introduce a buffer  gas in both sides of the experiment. By changing the refractive index of the medium, they were able to shift the dips in sensitivity due to the oscillatory nature of the probability (see equation~(\ref{prob})) and therefore fill those sensitivity gaps. In their second stage, planned in two steps, they expect to achieve the locking of the two resonant cavities and serious improvements in the laser power and the detector. We have estimated the discovery potential for the next generation of LSW experiments: figure~\ref{fig:bound1} displays the expected  sensitivity for ALPs and figure~{\ref{fig:bound2}}  the same for hidden photons. We have assumed the benchmark values   of table~\ref{table:t1}. 
As we can see, it might be possible that for the first time LSW experiments could be more sensitive than solar searches and scan new parameter space on ALPs . For hidden photons, they largely supersede the previous phase and in particular, the hidden CMB hypothesis (see section~\ref{sec3}) could be tested.
The sensitivity of these experiments has grown considerably over the last few years, to the point that by now they are
the most sensitive purely laboratory probes. {The advantage is that laboratory bounds are less model dependent, and they also apply
if the couplings to photons effectively depend on environmental conditions such as the temperature and matter\footnote{One can account for matter effects into the LSW conversion probability by introducing the polarization tensor into the equations of motion.}, providing a clean and controlled environment \cite{Jaeckel:2006xm}.

In this note we have reviewed the theoretical motivation for the existence of WISPs and some of their stronger hints, most of them coming from cosmology. We have then recapitulated one important search of WISPs with photons, the light shining through wall experiments and we have shown the major improvements so far proposed for the experiment and the expected sensitivity for the next generation.
\ack P. Arias acknowledges the valuable support of the Alexander von Humboldt Foundation.

\end{document}